# DISCIPLINARY KNOTS AND LEARNING PROBLEMS IN WAVES PHYSICS

**Simone Di Renzone**[1], **Serena Frati**[2]**, Vera Montalbano,** *Phys. Department, University of Siena*

**Abstract**
An investigation on student understanding of waves is performed during an optional laboratory realized in informal extracurricular way with few, interested and talented pupils. The background and smart intuitions of students rendered the learning path very dynamic and ambitious. The activities started by investigating the basic properties of waves by means of a Shive wave machine. In order to make quantitative observed phenomena, the students used a camcorder and series of measures were obtained from the captured images. By checking the resulting data, it arose some learning difficulties especially in activities related to the laboratory. This experience was the starting point for a further analysis on disciplinary knots and learning problems in the physics of waves in order to elaborate a teaching-learning proposal on this topic.

## 1. Introduction

Wave phenomena are everywhere. Everything can vibrate. There are oscillations and waves in water, ropes and springs. There are sound waves and electromagnetic waves. Even more important in physics is the wave phenomenon of quantum mechanics. When and how can it make sense to use the same word, wave, for all these disparate phenomena? What is it that they all have in common?

A first answer lies in the mathematics of wave phenomena. Periodic behaviour of any kind, one might argue, leads to similar mathematics. There is a more physical answer to the questions. If it is possible to recognize deep similarities in different physical phenomenology, then it is likely that we can describe them by mean of the same mathematical tools.

In order to introduce interested and talented students on this intriguing field in which wave phenomena can be described most insightfully, we designed an optional laboratory within National Plan for Science Degree[3].

In the following, we describe methodological choices made in planning an extracurricular learning path on waves and oscillations and some examples of activities. In particular, we outline the advantages of introducing mechanical waves by using the Shive wave machine (Shive 1959). Many laboratory activities can be proposed in which students explore waves behaviour in qualitative way, guess what can happen and suddenly test their hypothesis. Furthermore, we present same disciplinary knots that arise usually in empirical investigation, according to the Model of Educational Reconstruction (Duit 1997, 2007).

Finally, we describe learning problems on fundamental topics which arose in the laboratory. Further analysis on disciplinary knots is needed for a new teaching-learning proposal.

## 2. An extracurricular learning path on waves

In the last years, many Italian Universities are involved in National Plan for Science Degree.
The PLS guidelines are the following:
- orienting to Science Degree by means of training
- laboratory as a method not as a place
- student must become the main character of learning
- joint planning by teachers and university
- definition and focus on PLS laboratories.

There are several types of PLS laboratories. Laboratories which approach the discipline and develop vocations, self-assessment laboratories for improving the standard required by undergraduate courses and deepening laboratory for motivated and talented students.
We proposed two deepening laboratories for selected students[4] titled *Waves and energy* and *Sound and surroundings*. The laboratories are optional and the activities take place in Physics

---

[1] teacher enrolled in Master In Educational Innovation in Physics and Orienting - University of Udine
[2] teacher enrolled in Master In Educational Innovation in Physics and Orienting - University of Udine
[3] Piano nazionale Lauree Scientifiche, i.e. PLS.
[4] Coming from third class of *Liceo Scientifico Aldi* – Grosseto, followed by their teacher G. Gargani

Department. We are meeting students for 3 hours almost every month and planning to continue for last 3 years of high school. We decided that for the first year both laboratories had the same introductory activities on waves physics and all students works together.
We planned activities by focusing on
- conceptual issues such as characterization of the oscillatory motion and energy aspects vs. characterization of wave energy and energy transport
- methodological issues in order to propose a complementary experience compared to what was done in class.

Students ended their learning path on waves and sounds in class before the laboratory starts. Moreover, their class made an instruction trip to our department and perform a standard laboratory experience on diffraction and interference.

## 3. Shive wave machine

We chose to begin with a series of activities performed by student by mean of a Shive wave machine, showed in Fig. 1.a.

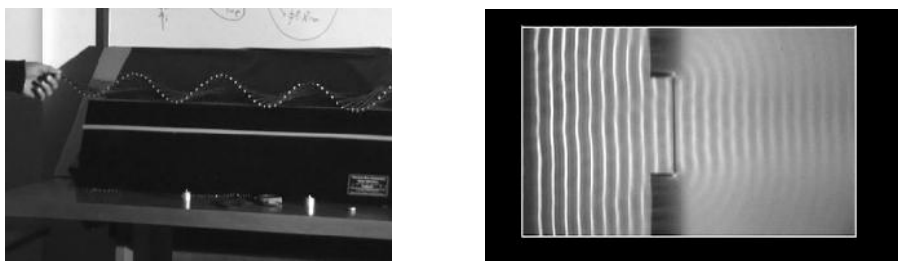

*Figure 1: a) Shive wave machine, b) wave tank*

This wave machine, developed by Dr John Shive at Bell Labs in '50, consists of a set of equally-spaced horizontal rods attached to a square wire spine. Displacing a rod on one of the ends will cause a wave to propagate across the machine. Torsion waves of the core wire translate into transverse waves.

Table 1 shows the main advantages and disadvantages of Shive wave machine vs. wave tank, another educational device very common in school.

*Table 1: Shive wave machine vs. wave tank*

| Shive Wave Machine | Wave Tank |
|---|---|
| Easy student interaction<br>Measure λ, T, v<br>Superposition principle<br>Reflection study<br>Energy considerations<br>Easy study of stationary waves and resonance | Surface waves<br>Measure λ, T, v<br>Reflection and refraction study<br>Interference and diffraction |
| *Limit of application* | |
| One-dimensional wave<br>Lack of study of refraction, diffraction and interference | Lack of information on energy<br>Poor interaction with student |

Measures were obtained by using a camcorder and extracted from the captured images.
Students used Shive wave machine for studying the following wave aspects:
- dependence on space-time of waves
- impulsive and periodic waves
- wavelength and frequency
- energy transfer

- speed of propagation
- superposition principle
- reflection and transmission

For example, by coupling core wires of two Shive wave machine with different rod lengths, an impulsive wave can be reflected and transmitted through the discontinuity (two examples are given in Fig.2). Students can measure from captured images all wave amplitudes and speeds and verify that energy is conserved.

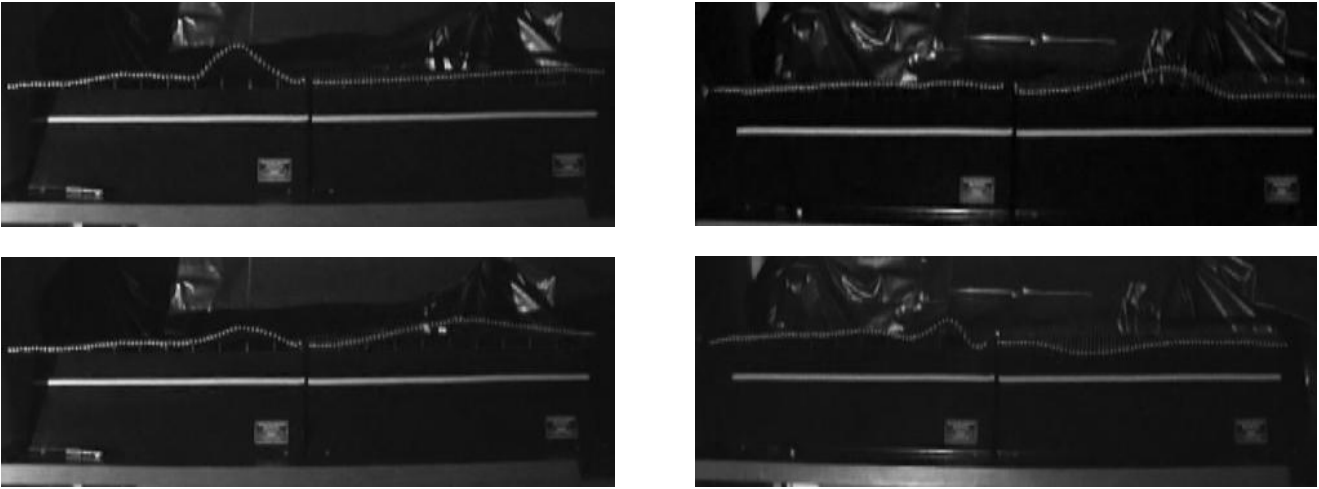

Figure 2: a) a pulse from the left (top) and transmitted and reflected pulses (below),

b) a pulse from the right (top) and transmitted and reflected pulses.

## 4. Further activities in laboratory

After using Shive wave machine, sound waves were studied by using a microphone and an oscilloscope (Fig. 3.a). Further verification of the principle of superimposition was made. Also beats and patterns of periodic beats (Moiré fringes) were studied .

Interference was studied for sound waves in order to stimulate reflection around similarities and differences between different kinds of waves (longitudinal vs. transverse, three-dimensional vs. two-dimensional, and so on). Fig. 3.b shows a schematic layout of interference experiment.

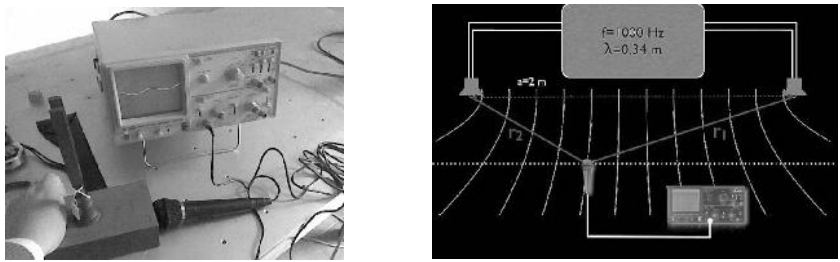

Figure 3: a) characterizing a sound waves by using an oscilloscope , b) acoustic interference

This experiment is complementary to interference with laser light and allows measuring the speed of sound and discussing problems and connected applications of interference in acoustics.

In order to study resonance, we started from a mechanical system with proper modes which can be forced by induction. A ceramic magnet is hanging to a string suspended over an electromagnet.

Students can vary the frequency of a square wave that power the electromagnet, they can measure the amplitude value varying the frequency of the forcing term.
The next steps will be to study resonance with the Shive wave machine, in acoustic and optical resonant cavities.

## 5. Some disciplinary knots in wave physics

Starting from our previous empirical investigation in teaching waves, we designed the learning path on waves described in previous sections in order to use it as a pilot study in an optional laboratory performed with few, interested and talented students, i. e. in the best conditions for teaching and deepening topics in physics.
We focused our attention on topics in which the main difficulties in learning usually appear.
We have considered the following conceptual knots:

- Waves as function of several variables; this usually is an hidden trouble. Even brilliant students can use for long times functions of one and several variables without any real understanding of deep difference.
- Energy transport is essential in order to distinguish waves from other periodic phenomena and comprehend many applications.
- Superposition principle is a fundamental concept. In wave physics, many phenomena can be clarified and some unexpected behavior can be explained by applying it.
- Analogy in waves phenomena; difficulties in this area are very common and reported (Podolefsky 2007a, b), especially in recognizing the same behaviour in different context such as total reflection, diffraction, beats, interference and so on.
- Resonance; despite it is a relevant phenomenon which runs through almost every branch of physics, many students have never studied it. Yet, resonance is one of the most striking and unexpected phenomenon in all physics and it easy to observe but difficult to understand.

## 6. Learning problems

Despite excellent boundary conditions (motivated and talented students), despite optimal behavior and relationship between instructors and pupils (we both have really fun in making this laboratory), we have encountered learning problems on fundamental topics.
The first learning problem arose when we requested graphics for Shive wave machine position of one rod vs. time (position fixed) and position of rods vs. space (time fixed), in order to outline the dependences from space and time variables and put them in relation with measurable physical quantities.
Home students' task was to extract all data from video captured by themselves. We had outlined the importance of calibrating the video for time and space measures by giving examples and hints.
Students prepared two graphics with 3 experimental points with no errors, the points were 0, max, 0, i.e. they measured only the maximum wave amplitude and the period or the wavelength, guessed the nodal point must be 0 and make graphics by drawing one half period, like shown in Fig. 4.

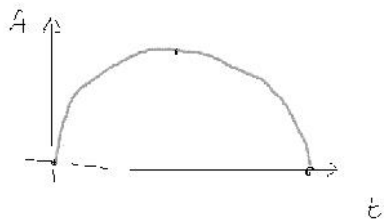

*Figure 4: A drawing like that shown was done on a large sheet of graph paper*

The next time students presented graphics with about seven points and uncertainties of about 15% or 20%
We were astonished. With camcorder, the data can have uncertainties of few percent.

After a brief discussion, we discovered that videos captured in the previous lab are useless for time calibration because digits on clock display are too small and they had forgot to measure a length in the devices that can be used as calibration.

Thus, they adopted the following strategy: measuring lengths on the computer display with a ruler and measuring time by a chronometer. You can imagine what precision could have measuring lengths on a computer display.....

On the other side, these students are very clever in executing tasks in which they have received instruction as in a laboratory at school where usually a worksheet is given.

In our opinion, they are still missing the main purpose of a measure: obtain the maximum of information from nature with the given devices.

Furthermore, how can they understand deeply the difference between a one dimensional wave and a function encountered in a math lesson, or between one dimensional wave and two-dimensional or space waves or surface waves, if they are not able to construct graphics that represent completely a physical situation starting from data?

## 7. Conclusions

We proposed and are testing a learning path in wave physics.

Some disciplinary knots were identified and laboratory activities were developed in order to help understanding in learning processes. Testing these activities in an optional laboratory with high school student can be considered the first step in order to develop a designed-based research learning path (Hake 2008).

These laboratories seemed to be very successful but in the meantime students showed serious lacking in fundamental topics, especially in collecting properly and in using correctly the experimental data in order to describe physical system.

These lacks are the basis of observed learning problems and remains a real trouble for any further progress in knowledge.

It is crucial to analyze deeply the teaching-learning processes in undergraduate school in order to avoid, or at least, minimize these kind of learning problems.

**References**

Duit R (2007), Science Education Research Internationally: Conceptions, Research Methods, Domain of Research, EJMSTE, 3, 3-15

Duit R., Komerek M., Wilbers J. (1997), Studies on Educational Reconstruction of Chaos Theory, Research in Science Education, 27 (3), 339-357

Hake, R.R. (2008), Design-Based Research in Physics Education Research: A Review," in Kelly, Lesh, & Baek, 493-508

Podolefsky N. S., Finkelstein N. D. (2007a), Salience of Representations and Analogies in Physics, AIP Conf. Proc. 951, pp. 164-167, 2007 PHYSICS EDUCATION RESEARCH CONFERENCE

Podolefsky N. S., Finkelstein N. D. (2007b), Analogical scaffolding and the learning of abstract ideas in physics: An example from eletromagnetic waves, *Phys. Rev. ST - Phys. Educ. Res.* 3, 010109

Shive J. N. (1959), Similarities of Wave Behavior, http://techchannel.att.com/play-video.cfm/2011/3/7/AT&T-Archives-Similarities-of-Wave-Behavior, accessed 2011 October

.